\documentclass[12pt]{article}
\usepackage{amsmath,amsfonts,amssymb,graphicx,multirow}
\usepackage{graphicx}
\usepackage{enumerate}
\usepackage{natbib}
\usepackage{url} % not crucial - just used below for the URL 
\usepackage{etoolbox}

\newcommand{\blind}{0}

\newcommand{\argmax}{\mathop{\mbox{argmax}}}
\newcommand{\var}{\mathop{\mbox{var}}}

\newcommand{\cosk}{\mathop{\mbox{cosk}}}

\makeatletter

\patchcmd{\NAT@test}{\else \NAT@nm}{\else \NAT@nmfmt{\NAT@nm}}{}{}

\DeclareRobustCommand\citepos
{\begingroup
	\let\NAT@nmfmt\NAT@posfmt% ...except with a different name format
	\NAT@swafalse\let\NAT@ctype\z@\NAT@partrue
	\@ifstar{\NAT@fulltrue\NAT@citetp}{\NAT@fullfalse\NAT@citetp}}

\let\NAT@orig@nmfmt\NAT@nmfmt
\def\NAT@posfmt#1{\NAT@orig@nmfmt{#1's}}

\makeatother

\begin{document}

% \nocite{*}
%\bibliographystyle{natbib}

\def\spacingset#1{\renewcommand{\baselinestretch}%
{#1}\small\normalsize} \spacingset{1}

%%%%%%%%%%%%%%%%%%%%%%%%%%%%%%%%%%%%%%%%%%%%%%%%%%%%%%%%%%%%%%%%%%%%%%%%%%%%%%

\if0\blind
{
  \title{\bf Projection pursuit based generalized betas accounting for higher order co-moment effects in financial market analysis}
  \author{Sven Serneels \\ Aspen Technology, Bedford, MA, USA}
  \maketitle
} \fi

\if1\blind
{
  \bigskip
  \bigskip
  \bigskip
  \begin{center}
    {\LARGE\bf Projection pursuit revisited as a flexible framework for dimensionality reduction}
\end{center}
  \medskip
} \fi
%The text of your abstract.  200 or fewer words.
\bigskip
\begin{abstract}

Betas are possibly the most frequently applied tool to analyze how securities relate to the market. While in very widespread use, betas only express dynamics derived from second moment statistics. Financial returns data often deviate from normal assumptions in the sense that they have significant third and fourth order moments and contain outliers. This paper targets to introduce a way to calculate generalized betas that also account for higher order moment effects, while maintaining the conceptual simplicity and interpretability of betas. Thereunto, the co-moment analysis projection index (CAPI) is introduced. When applied as a projection index in the projection pursuit (PP) framework, generalized betas are obtained as the directions optimizing the CAPI objective. A version of CAPI based on trimmed means is introduced as well, which is more stable in the presence of outliers. Simulation results underpin the statistical properties of all projections and a small, yet highly illustrative example is presented.  

\end{abstract}

\noindent%
\vfill

\newpage
\spacingset{1.5} % DON'T change the spacing!

\section{Introduction}
\label{sec:intro}

In financial market analysis, it is of great interest to know how each security relates to a reference index. The most common reference index would be {\em the market}. The latter can be represented by generally accepted market indices. For the US markets, good examples of the reference index would be the Dow Jones Industrial Average or the Standard and Poor's 500 index. 

Interpreting how each security relates to the market is very commonplace practice. The most widespread way to do this is through the {\em betas}: the regression coefficients of the returns of the corresponding security to the returns of the market. The signs of the betas indicate which security correlates positively or negatively with the market, and the magnitude of the betas indicate how volatile each security is compared to the market. Betas are supplied by default as a part of the description of each stock in most brokerage software, as well as in financial portals such as Yahoo Finance (\url{http://finance.yahoo.com}). While betas convey very valuable and actionable information, they do not portray the full picture to understand how securities relate to the market. They are regression coefficients of individual securities to the market (or a sector thereof) and are thus based on first and second order moment statistics. When normal distribution assumptions hold true, such statistics are sufficient. However, financial returns data often deviate from the normal distribution in the sense that they have significant higher order moments and may contain outliers. Most traditional (and widely applied) market analysis techniques will silently assume that (log)returns are normally distributed, while in reality, they often have significantly nonzero skewnesses and kurtoses. Also, the multivariate effect cannot be ignored: co-skewness and co-kurtosis are typically significant. 

The target of this paper is to introduce a measure of market dynamics that is as conceptually straightforward as betas, yet accounts for higher order moment effects. Eventually, this paper introduces generalized betas as the directions optimizing the co-moment analysis projection index (CAPI) within the Projection Pursuit (PP) framework. These generalized betas essentially express all relevant market dynamics as one single number and the weight attributed to the contribution of higher order moments can be configured based on investor preference. The concept of generalizing betas to higher moments has been raised before by \cite{Martinelli2007}, who come up with individual definitions specific generalized betas for each higher order moment. Besides the necessity to calculate a separate generalized beta for each moment, these betas also have to be calculated independently for each security analyzed. The novelty of the method introduced in this paper resides in the fact that the resulting generalized betas are one single number summarizing all relevant second and higher order moment effects and that all betas are obtained simultaneously as the result of one single calculation. The latter is owed to the generalized betas being the result of a projection pursuit algorithm. 

While the methods proposed are eventually highly practical, the paper will start out by focusing on the concept of projection pursuit itself. In Section \ref{Sec:PI}, different viable higher order moment projection indices will be discussed, and the CAPI projection index, which leads to generalized betas, will be introduced. Section \ref{Sec:Algorithms} elaborates on algorithms to calculate projection pursuit solutions. Then, a simulation study underpins that projection pursuit does retrieve solutions with expected properties. Section \ref{Sec:Application} presents a small, yet highly illustrative example, whereas Section \ref{Sec:Conclusions} concludes. 

\section{Projection Pursuit \label{sec:PP}}

Projection pursuit (PP) has a long standing history as a a tool to extract relevant subspaces from high-dimensional spaces. The concept dates back to \citet{Kruskal1969} and then been extensively investigated by \citet{FriedmanTukey1974} as a tool for exploratory data analysis, who also coined its catchy name. Projection pursuit is essentially a model free dimension reduction technique: it consists of finding a set of direction vectors $\mathbf{a}$ satisfying optimality according to a given optimization criterion under a predefined set of constraints.  

Let $\mathbf{x}$ denote a $p$-variate random vector with associated probability distribution $F(\mathbf{x}): \mathbb{R}^p \rightarrow \mathbb{R}$ and denote by $\mathfrak{F}$ the space of probability distribution functions. Projection pursuit accomplishes dimensionality reduction by scanning all $p$-variate vectors $\mathbf{a}$ such that the linear projections $\mathbf{a}^T\mathbf{x}$ satisfy optimality according to a certain criterion. Since $\mathbf{x} \sim F$, the linear projection $\mathbf{a}^T\mathbf{x}$ is a univariate random variable with distribution function $F_{\mathbf{a}}$. Attaining projection pursuit optimality, corresponds optimizing a functional $\mathcal{P}(F_{\mathbf{a}}): \mathfrak{F} \rightarrow \mathbb{R}$: 
\begin{equation}\label{eq:ppdef_func}
     \mathbf{w} = \argmax_{\mathbf{a}} \mathcal{P}\left(\mathbf{a}^T\mathbf{x},F_{\mathbf{a}}\right).
\end{equation}
The functional $\mathcal{P}$ can take on any shape. Various options leading to sensible statistical dimensionality reduction will be discussed in section \ref{Sec:PI}. For instance, a common choice for $\mathcal{P}$ is setting it equal to variance. Let $u$ be a univariate random variable with (differentiable) distribution $F_u$. Given that $F_u$ has an existing first moment, $u$ can without loss of generality be assumed to be centred, i.e. $E(u,F_u) = \int u dF_u = 0$. For the sake of simplicity, this assumption will be made throughout the article. Then the variance of $u$ equals the second moment of $F_u$: 
\begin{equation}\label{eq:var_func}
 \mathcal{P}_{\var{}} = \var{(u,F_u)} = \int {u^2 dF_u}. 
\end{equation}
In practice, one will only have a finite sample $\mathbf{X}$, a matrix consisting of $n$ cases of $\mathbf{x}$. Based on such a finite sample, the functionals $\mathcal{P}$ will now be evaluated at the empirical distribution $F_{n,\mathbf{a}}$. For the variance example, this corresponds to 
\begin{equation}\label{eq:ppdef_empir}
     \hat{\mathbf{w}}_1 = \argmax_{\mathbf{a}} \mathcal{P}_{\var{}} \left(\left(\mathbf{a}^T\mathbf{x},F_{n,\mathbf{a}}\right)\right) = \argmax_{\mathbf{a}} \left(\mathbf{a}^T\mathbf{X}\mathbf{X}^T\mathbf{a}\right). 
\end{equation}
With a unity norm constraint for $\mathbf{w}$, this corresponds to finding the first principal component in principal component analysis (PCA). 

As soon as a first optimal direction $\hat{\mathbf{w}}_1$ has been obtained, ensuing optimal directions can be identified by repeating the procedure in the $(p-h+1)$-dimensional subspace orthogonal to subspace spanned by the vectors that have already been estimated. This can be achieved by defining a constrained optimization: 
\begin{subequations}\label{eq:ppdef_funcc}
\begin{equation}\label{eq:ppdef_ctric}
\mathbf{w}_i = \argmax_{\mathbf{a}} \mathcal{P}\left(\mathbf{a}^T\mathbf{x},F_{\mathbf{a}}\right), 
\end{equation}
subject to:
\begin{equation}\label{eq:ppdef_constrc}
\mathbf{w}_i^T E\left(\mathbf{x}\mathbf{x}^T\right)\mathbf{w}_j = 0 \mbox{ and } \parallel \mathbf{w}_i\parallel = 1, 
\end{equation}
\end{subequations}
where $\parallel \cdot \parallel$ denotes the Euclidean norm and $j < i \in [1,p]$.
Setting $\mathcal{P} = \mathcal{P}_{\var{}}$ in \eqref{eq:ppdef_ctric} is now integrally equivalent to principal component analysis. 

Because of its flexibility, projection pursuit has found many applications of drastically different nature. By 1985, projection pursuit had found its way into new metods for (non-linear) regression, density approximation, density estimation and even time series problems. An excellent overview of these methods has been provided by \citet{Huber1985}. 

Projection pursuit is a very general method and can either be applied directly for dimensionality reduction, where analyzing the components $\mathbf{w}$ would be of interest, or as a plugin to a more complex construct, such as PP regression. In the former context, PP has mainly become popular in the area of robust statistics. Whereas PCA technically is a PP based dimension reduction technique, in practice, direct PP algorithms scanning all directions $\mathbf{a}$ will typically not be applied to calculate principal components, since the PCA solution to Criterion \eqref{eq:ppdef_funcc} can be obtained analytically and the PP solution can thus be calculated more efficiently, e.g. through eigenvalue decomposition. Direct projection pursuit algorithms get their practical use in those situations where no analytical solution to the optimization criterion can be obtained. Robust dimension reduction techniques constitute a good example of this situation, since they typically involve maximizing a robust or nonparametric projection index that may be very involved or impossible to optimize analytically. In this context, projection pursuit has been investigated quite a bit. Research on robustified dimensionality reduction through projection pursuit started with robust principal component analysis, the earliest version of which goes back to \cite{ChenLi1981}, later also published more generally in \citep{LiChen1985}. These authors construct a robust PCA method by using the median absolute deviation as a projection index, which is widely known to be a robust measure of scale. \cite{Croux:Proj} revisit the same approach, using the more statistically efficient $q_n$ estimator of scale \citep{RousseeuwCroux1993} as a projection index. Several algorithmic aspects have been developed since, which will be discussed in Section \ref{Sec:Algorithms}. More recently, a sparse and robust alternative to PCA has been developed based on PP as well \citep{CrouxFritzFilzmoser2013}.

Whereas application of (classical and robust) scale based projection indices has extensively been explored, the field becomes much more sparse for projection indices involving a dependent variable. The present paper will focus on methods resulting from such projection indices, a selection of which will be highlighted in Section \ref{Sec:PI}. More specifically, classical and robust projection indices will be deployed that enable analyzing nongaussianity in the the relationship to a dependent variable. The resulting dimensionality reduction methods can be of great practical utility in financial data analysis, but also elsewhere.

\section{Projection Indices}\label{Sec:PI}

This Section will focus on projection indices that involve a dependent variable. In fact, it is straightforward to generalize Equations \eqref{eq:ppdef_funcc} to a setting where there is a connection to a dependent variable. When the objective is to analyze securities, the dependent variable typically is a market index or a sector index. However, what follows is much more generally applicable and is therefore presented as such. 

Let $y$ be a univariate random variable with probability distribution $F_y$. Typically, it will be assumed that there exists a bivariate distribution $F_2$ of which both $F_y$ and $F_{\mathbf{a}}$ are marginals. Then components that achieve dimension reduction involving both random variables, can be defined as: 
\begin{subequations}\label{eq:ppdef_funccy}
	\begin{equation}\label{eq:ppdef_ctricy}
	\mathbf{w}_i = \argmax_{\mathbf{a}} \mathcal{P}\left(\mathbf{a}^T\mathbf{x},y,F_2\right), 
	\end{equation}
	subject to:
	\begin{equation}\label{eq:ppdef_constrcy}
	\mathbf{w}_i^T E\left(\mathbf{x}\mathbf{x}^T\right)\mathbf{w}_j = 0 \mbox{ and } \parallel \mathbf{w}_i\parallel = 1, 
	\end{equation}
\end{subequations}
where now the projection index is a functional of both random variables. Here, the most straightforward example is covariance: 
\begin{equation}\label{eq:cov_func}
\mathcal{P}_{\mathrm{cov}} = \mathcal{M}_2(\cdot,\cdot) =  \mathrm{cov}{(u,v,F_{u,v})} = \int{uv\ dF_{u,v}}, 
\end{equation}
where $u$ and $v$ are variables that are drawn from a bivariate distribution $F_{u,v}$. It is clear that Equation \eqref{eq:cov_func} represents the second order co-moment.  Depending on the target of the analysis, it may be preferable to plug in the square of this measure instead. 
Using squared covariance $\mathcal{P}_{\mathrm{cov}}^2$ from \eqref{eq:cov_func} as a projection index in \eqref{eq:ppdef_ctricy} actually defines partial least squares (PLS) on a population level. Again, direct projection pursuit algorithms calculating a solution by numerically evaluating many directions $\mathbf{a}$, are not used to calculate the partial least squares solution, since in the special case of covariance, Criterion \eqref{eq:ppdef_ctricy} can be solved analytically, leading to more computationally efficient implementation options. Also here, direct projection pursuit is applied to estimate robust partial least squares. That approach has been proposed by \cite{SFCV2005} as a specific case of the more general robust continuum regression (RCR) method. In particular, robust PLS is obtained when setting RCR's parameter $\delta$ to $\delta=\frac{1}{2}$.

For each co-moment projection index, it is possible to define a scaled and a non-scaled version. Let $\sigma_u$ and $\sigma_v$ denote the marginal scale parameters of $F_{u,v}$. Then the scaled second order co-moment is:
\begin{equation}\label{eq:corr_func}
\mathcal{P}_{\mathrm{corr}} = \acute{\mathcal{M}}_2(u,v,F_{u,v}) =  \mathrm{corr}{(u,v,F_{u,v})} = \int{\frac{uv}{\sigma_u\sigma_v}dF_{u,v}}, 
\end{equation}
which obviously corresponds to population correlation. Here and elsewhere, the acute accent ($\acute{ }$) will denote the scaled version of a co-moment estimator. Plugging the correlation projection index from \eqref{eq:corr_func} into criterion \eqref{eq:ppdef_ctricy} defines Canonical Correlation Analysis (CCA). 

While projection indices based on second order co-moments have attracted a fair share of attention in the literature, there is little to no literature on construction of directions that achieve optimality according to projection indices based on higher order co-moments. In this paper, we fill mainly focus on the applicability of co-skewness and co-kurtosis as projection indices.
Co-skewness between two (univariate) random variables $\mathbf{u}$ and $\mathbf{v}$ is not uniquely defined, but there exist two measures of co-skewness: 
\begin{subequations}\label{eq:cosk_func}
\begin{equation}\label{eq:cosk0_func}
\mathcal{P}_{\mathrm{cosk1}} = \acute{\mathcal{M}}_{3,1}(u,v,F_{u,v}) = \int {\frac{u^2v}{\sigma_u^2\sigma_v} dF_{u,v}}
\end{equation}
and
\begin{equation}\label{eq:cosk1_func}
\mathcal{P}_{\mathrm{cosk2}} = \acute{\mathcal{M}}_{3,2}(u,v,F_{u,v}) = \int {\frac{uv^2}{\sigma_u\sigma_v^2} dF_{u,v}}.
\end{equation}
\end{subequations}
The coresponding non-scaled co-moments $\mathcal{M}_{3,j}, j \in [1,2] \subset \mathbb{N}$, have a denominator of 1 in the integrals in Equations \eqref{eq:cosk_func}.  
Note that both measures of co-skewness are elements of the bivariate distribution's co-skewness matrix (often denoted $\boldsymbol{\Phi}$). 

For finite samples, the co-skewness can be calculated as 
\begin{equation}\label{eq:cosk0_sample}
\mathrm{cosk1}{(\mathbf{u},\mathbf{v},F_{n_{u,v}})} = c \sum_i u_i^2 v_i / {\sigma_u^2\sigma_v},
\end{equation}
where $c$ is generally set to 1, but for finite samples at the normal distribution, can be set to the consistency factor $c = n / ((n-1)(n-2))$. 

Likewise, co-kurtosis between two (univariate) random variables $\mathbf{u}$ and $\mathbf{v}$ comes in three versions: 
\begin{subequations}\label{eq:coku_func}
	\begin{equation}\label{eq:cok0_func}
	\mathcal{P}_{\mathrm{coku1}} = \acute{\mathcal{M}}_{4,1}(u,v,F_{u,v}) = \int {\frac{u^3v}{\sigma_u^3\sigma_v} dF_{u,v}},
	\end{equation}
	\begin{equation}\label{eq:cok1_func}
	\mathcal{P}_{\mathrm{coku2}} = \acute{\mathcal{M}}_{4,2}(u,v,F_{u,v}) = \int {\frac{u^2v^2}{\sigma_u^2\sigma_v^2} dF_{u,v}}
	\end{equation}
	and
	\begin{equation}\label{eq:cok2_func}
	\mathcal{P}_{\mathrm{coku3}} = \acute{\mathcal{M}}_{4,3}(u,v,F_{u,v}) = \int {\frac{uv^3}{\sigma_u\sigma_v^3} dF_{u,v}}
	\end{equation}
\end{subequations}
Also here, the corresponding non-scaled fourth order co-moments $\mathcal{M}_{4,\ell}(\cdot,\cdot,\cdot,\cdot), \ell \in [1,3]\subset \mathbb{N}$ can be defined  similarly by setting the denominator to 1 in the integrals in Equations \eqref{eq:coku_func}. 

Finally, it is possible to construct composite projection indices that are combinations of different moments. As such, the CAPI projection index is introduced: the co-moment analysis projection index. It is defined as: 
\begin{equation}\label{eq:capi}
\mathcal{P}_{\mathop{\mbox{CAPI}}} = \sum_{ij} \omega_{ij} \mathcal{M}_{ij},
\end{equation}
where $i \in [1,4]\subset \mathbb{N}$ loops through the orders and $j$ loops trough the available versions of the co-moment, i.e. $j \in [1,i-1] \subset \mathbb{N}$. Likewise, a scaled version $\acute{\mathcal{P}}_{\mathop{\mbox{CAPI}}}$ of CAPI can be defined:
\begin{equation}\label{eq:capiscaled}
\acute{\mathcal{P}}_{\mathop{\mbox{CAPI}}} = \sum_{ij} \omega_{ij} \acute{\mathcal{M}}_{ij},
\end{equation} wherein now the $\mathcal{M}_{ij}$ have been substituted by correlation, co-skewness and co-kurtosis, respectively. 

Essentially, CAPI is a linear combination of co-moments up to a certain order, wherein the $\omega_{ij}$ represent the weight attributed to each co-moment.

\section{Algorithms}\label{Sec:Algorithms}

Any algorithm for direct projection pursuit essentially consists of two components: a section to find the direction $\mathbf{a}$ optimizing Criterion \eqref{eq:ppdef_ctricy}, and a projection based algorithmic construction to ascertain that the side constraints in \eqref{eq:ppdef_constrcy} are being respected. Both steps are common aspects of direct projection pursuit algorithms, regardless of the projection index and regardless of the presence of a dependent variable. Therefore, existing algorithms from the literature can be adopted here. 

The oldest approach to direct projection pursuit, is to randomly generate a candidate set of $p$-variate vectors $\mathbf{a}$ and then for each of these, to evaluate the projection index $\mathcal{P}$. This approach was, among other papers, adopted in \citepos{LiChen1985} approach to robust PCA. An issue with this approach can obviously be that is is not straightforward to find the right candidate set. Therefore, the outcome of the algorithm may be unstable, unless if a very large number of directions is being selected. Selecting a very large candidate set may lead to lengthy calculation times. \citet{Croux:Proj_orig} improve on this idea by selecting a candidate set that includes that actual training data. The rationale behind doing so, is that it is more likely to find the optimum {\em in the direction of the data} and therefore, the algorithm should converge with fewer directions and should be more stable. A very similar algorithm to \citet{Croux:Proj_orig} was developed by \citet{RAPCA} and baptised {\em reflection-based algorithm for principal component analysis} (RAPCA). \citet{RAPCA} report the algorithm in \citet{Croux:Proj_orig} to be unstable and conjecture that to be due to a sequence of round-off errors that deteriorates the results as $p$ increases. However, more recently, \citet{CrouxFilzmoserOliveira2008} have proven that both of these algorithms are fundamentally flawed, since both of them suffer from the {\em m exact fit} error, which means that they will essentially estimate directions with zero scale due to absence of a sufficient amount of data when correcting for outliers in high dimensions. \citet{CrouxFilzmoserOliveira2008} mention that these flaws are remedied in the {\em grid} algorithm. The grid algorithm had in fact been developed a few years earlier by \citet{FSCV2006} as a faster and more stable algorithm for robust continuum regression \citep{SFCV2005}, the original algorithm of which had suffered from the same stability flaws as the corresponding PP algorithms for robust PCA. \citet{FSCV2006} report the grid algorithm to outperform earlier approaches, without reporting theoretical justification, which was then provided by \citet{CrouxFilzmoserOliveira2008}. Since its inception, the grid algorithm has gained widespread traction. It has, for instance, been included in expectation---robust (ER) routines that allow for model consistent missing value imputation while estimating robust PCA \citep{SerneelsVerdonck2008} and principal component regression \citep{SerneelsVerdonck2009} models. It has also been included in the \textsf{R} packages \textsf{rrcov} \citep{TodorovFilzmoser2009} and \textsf{rrcovNA} for robust multivariate analysis, the latter of which incorporates the aforementioned ER approaches for missing value imputation.  

In brief, the grid algorithm proceeds by subdividing space into sections of the unit circle for each variable separately. Directions $\mathbf{a}$ are constructed as all combinations of these variable-wise subdivisions along the unit circle. Then $\mathcal{P}$ is evaluated for each direction. The combination of univariate subdivisions yielding optimal value evaluating $\mathcal{P}$, is retained. Then the subspace in which the optimum has been identified, is subdivided on a variable wise basis into subsegments on its turn. The optimum is identified in the new, smaller subspace. Subdivision of the optimal subspace is continued until convergence is met on the resulting $\hat{\mathbf{w}}$. This eventually satisfies \eqref{eq:ppdef_ctricy}. Details on how to implement this algorithm, can be found in \citet{FSCV2006}. 

The grid algorithm is applied so as to yield the optimal direction $\hat{\mathbf{w}}_i$. This direction is sometimes called a {\em weighting vector}, since it attributes weights to the input variables. However, at this point, it is only clear how to calculate $\hat{\mathbf{w}}_1$. The orthogonality constraint \eqref{eq:ppdef_constrcy} is now guaranteed by dint of a deflation scheme, that goes as follows. Let $\mathbf{E}_0 = \mathbf{X}$ be the training data. Then, for $i \in [1,\min(n,p)]$, recursively calculate: 
\begin{subequations}\label{eq:deflations}
	\begin{equation}\label{eq:weights}
	\hat{\mathbf{w}}_i = \argmax_{\mathbf{a}} \mathcal{P} \left(\left(\mathbf{a}^T\mathbf{E}_{i-1},\mathbf{y},F_{n,2}\right)\right)
	\end{equation}
	\begin{equation}\label{eq:scores}
	\hat{\mathbf{t}}_i = \mathbf{E}_{i-1}\hat{\mathbf{w}}_i
	\end{equation}
	\begin{equation}\label{eq:loadings}
	\hat{\mathbf{p}}_i = \mathbf{E}_{i-1}^T\hat{\mathbf{t}}_i / \left(\parallel\hat{\mathbf{t}}_i\parallel^2\right)
	\end{equation}
	\begin{equation}\label{eq:deflatedX}
	\mathbf{E}_i = \mathbf{E}_{i-1} - \hat{\mathbf{t}}_i \hat{\mathbf{p}}_i^T
	\end{equation}	
	
\end{subequations}   
Here, $F_{n,2}$ denotes the empiraical distribution corresponding to $F_2$. The maximization in \eqref{eq:weights} is done by applying the grid algorithm. Usually, the $\hat{\mathbf{t}}_i$ are called the {\em scores} or {\em components} and $\hat{\mathbf{p}}_i$ the {\em loadings}. We note though, that in independent component analysis (ICA), some variants of which also can be computed by this algorithm, typically a different nomenclature is used. At this point, it is noteworthy to state that, once the estimates of these components have been obtained, it is also possible to estimate a regression relationship between the scores and the dependent variable. Predictions for the dependent variable can be obtained by: 
	\begin{equation}\label{eq:yhatreg}
	\hat{\mathbf{y}} = \hat{\mathbf{T}} \hat{\boldsymbol{\gamma}},
	\end{equation}	
where the score matrix $\hat{\mathbf{T}}$ contains the scores $\hat{\mathbf{t}}_i$ in its columns and $\hat{\boldsymbol{\gamma}}$ is an estimate of regression coefficients of the relationship: 
\begin{equation}\label{eq:gammareg}
\hat{\boldsymbol{\gamma}} = \mathbf{y} \backslash \hat{\mathbf{T}},
\end{equation}
where $\backslash$ denotes estimation of a regression relationship, which in the case of least squares, amounts to: 
\begin{equation}\label{eq:gammareg}
\hat{\boldsymbol{\gamma}} = \hat{\mathbf{T}}\hat{\mathbf{T}}^T\mathbf{y},
\end{equation}
since the scores are orthogonal due to \eqref{eq:ppdef_constrcy}. 

Least squares is a common choice for estimation of $\boldsymbol{\gamma}$ estimating covariance relationships and is exactly how this {\em inner relationship} is estimated in PCR and PLS. However, when analyzing higher order statistics, it may be more informative to apply quantile regression instead of least squares. Also, in the presence of outliers, a robust regression estimate should be used, since the estimate in \eqref{eq:gammareg} can still be distorted by outliers. For that reason, RCR uses a robust M-regression (RM) estimate at this step \citep{SFCV2005}.

\section{Simulation}\label{Sec:Simulation}

The intent of this simulation study is to focus on co-skewness effects in the projection index. Performance of PP methods based classical second moment projection indices is well known. The performance of robust alternatives to these second order product---moment based methods has been well documented as well and need not be investigated here. However, no literature is available that investigates third order product---moment based projection indices, or robustified alternatives thereof. 

The targets of this simulation study are twofold. At first, the target of this study is to establish that PP is a suitable method to analyze common, yet indirectly observed sources of co-skewness in data. Secondly, it is well known that product---moment based estimators are very sensitive to outliers. For that reason, univariate robust alternatives have been developed for skewness, such as the Repeated Medtriple (RMT) and the Medcouple \citep{BrysStruyf2003,BrysStruyf2004,HalbertWhite2004}. Robustification of third order product---co-moment based estimators such as co-skewness is warranted, yet less well documented. As mentioned earlier, the approach pursued here is to use $\alpha$\% trimmed means in the internal calculation of these product moments, yielding a robust estimate. The target of this simulation study is to establish that PP based projections still capture the information in the data relevant to explain co-skewness in the presence of outliers. 

The assumption that a common latent, unobservable factor influences skewness, is made elsewhere in the literature as well. Particularly in the context of analyzing returns on financial assets, this is a viable assumption, but its application is not limited to that context. This assumption has recently even been included in a fast and precise way to estimate the skewness of linear combinations of random variables through a more precise estimate of the co-skewness matrix (\cite{Boudt2018}).  

In this study, however, we do not only target to estimate components with a common co-skewness within $\mathbf{X}$, but actually with maximal co-skewness to an exogenous variable $\mathbf{y}$. In order to construct such data, the latent variable $\mathbf{t}$ and the exogenous variable $\mathbf{y}$ are generated simultaneously from a bivariate skew $t$ distribution. Skew normal, skew $t$ and skew Cauchy distributions have been described in \cite{Azzalini2014} and the \textsf{R} package \textsf{sn} has been distributed by the same Author. Therein, these distributions have been implemented, as well as algorithms to draw random samples from them.  

In these skew-$t$ distributions, one can specify four parameters: 
\begin{itemize}
	\item $\boldsymbol{\xi}$, location parameter, 
	\item $\boldsymbol{\Omega}$, covariance matrix,
	\item $\boldsymbol{\alpha}$, the {\em slant} and
	\item $\nu$, the degrees of freedom.    
\end{itemize}   
Unlike simulating multivariate normal data, these parameters are not guaranteed to be the exact parameters of the distributions of the simulated data, yet parameters of the empirical distribution of simulated data will be close to the set parameters. For instance, data simulated from a skew $t$ distribution with $\boldsymbol{\Omega}=\mathbf{I}$, with $\mathbf{I}$ the identity matrix, will be approximately uncorrelated. 

Thus, in each simulation run, true $\mathbf{t}$ and $\mathbf{y}$ are generated as
\begin{equation}
  [\mathbf{y} \  \mathbf{t}] \sim t_{\mbox{skew}}(\mathbf{0},\boldsymbol{\Omega},\boldsymbol{\alpha},\nu),
\end{equation}
where in each simulation run a sample of size $n=1000$ is drawn. The degrees of freedom considered in this simulation study are $\nu \in \{1,2,5,50\}$. This gives access to skew-$t$ distributions ranging from a skew Cauchy to an almost skew normal. Both settings leading to uncorrelated $[\mathbf{y} \  \mathbf{t}]$ ($\boldsymbol{\Omega}=\mathbf{I}$), as well as correlated, are investigated. The latter are obtained by setting the off-diagonal elements of $\boldsymbol{\Omega}$ to 0.6. 
Then $\mathbf{X}$ is generated from $\mathbf{t}$ according to a classical latent variable model:  
\begin{equation}\label{eq:lvmodel}
\mathbf{X} = \mathbf{t}\mathbf{p}^T + [\epsilon_{ij}],
\end{equation}
where the $\epsilon_{ij} \sim N(0,\sigma_{\epsilon})$ is normally distributed noise with a certain preset scale. The most informative settings to investigate are those data have a nonzero, yet proportionally low noise level. Here, results for $\sigma_{\epsilon}=0.001$ are reported, as well as some results for $\sigma_{\epsilon}=1$. There is no theoretical limitation on what numerical values $\mathbf{p}$ can assume, nor on the amount of dimensions. However, some choices for $\mathbf{p}$ are more interesting than other ones. In this study, we opt to set $\mathbf{p} = [1\ 2\ 0.5\ 0.003\ 1.5]$. This yields five-dimensional data, the moderate dimensionality of which is still amenable to interpretation. Moreover, the small value for $p_4$ should be reflected in a small value for the estimated loading for this variable in the models.  

On top of varying the setting described above, the simulations are both run as-is, that is with data exactly described as above, and with contaminated data. Contamination is implemented as introducing $\phi = 10 \%$ of harsh outliers. Two ways to contaminate the data are investigated. At first, we look into the effect of contaminated {\em latent} and dependent variables i.e. by drawing both the first 5\% of the $y_i$ and the final 5\% of the $t_i$ from $N(25,1)$. Secondly, results are also reported for outliers in the $mathbf{X}$ data from \eqref{eq:lvmodel}. These are generated by subsituting the first 5\% of $\mathbf{x}_1$ and the final 5\% of $\mathbf{x}_2$ from $N(25,1)$.  

The results are structured as follows. Across $N=100$  simulations per setting, the locations and scales of the absolute deviations
\begin{equation}
\mathop{\mbox{AD}} = \mathop{\mbox{abs}}\left(\cosk\left(\mathbf{y},\mathbf{y},\hat{\mathbf{t}}_1\right)\right) - \left(\cosk\left(\mathbf{y},\mathbf{y},{\mathbf{t}}_1\right)\right),
\end{equation} are presented, where the estimates $\hat{\mathbf{t}}_1$ have been obtained by PP with classical or 15\% trimmed co-skewness as a projection index. These results are presented in Table \ref{tab:simres_cosk_u}. Then also the loading ratios, i.e. 
\begin{equation}
\mathop{\mbox{PR}} = \hat{p}_4 / \hat{p}_2
\end{equation}
are presented in Table \ref{tab:simres_pr_u}. 
   
\begin{table}
	\caption{\label{tab:simres_cosk_u} Location of Absolute Deviations (AD) of estimated $\cosk(\mathbf{y},\mathbf{y},\hat{\mathbf{t}}_1)$ to true for PP (co-skewness) estimates with varying degree of trimming. Numbers between brackets represent scale estimates (standard deviation for mean, MAD for median).}
	\begin{center}
	\begin{tabular}{ccccccc}
	   \hline
	   $\sigma_{\epsilon}$ & $\nu$ & $\phi$ & $\Omega_{ij}$ &
       Mean AD, PP(0) & 
       Median AD, PP(0) &
       Mean AD, PP(15) \\
       \hline
       0.001 &	1 & 0 & 0 &	3.13 (7.21) & 0.02 (4.64) &	11.66 (9.67) \\
       1 &	1 & 0 & 0 &	6.22 (9.61) & 0.15 (7.64) & 11.82 (9.67) \\
       0.001 &	2 & 0 & 0 &	1.95 (4.62) & 0.07 (2.76) &	4.55 (6.38) \\
       1 &	2 & 0 & 0 &	1.39 (3.54) & 0.09 (1.82) &	3.40 (4.57) \\
       0.001 & 5 & 0 & 0 & 0.09 (0.14) & 0.04 (0.08) &	0.25 (0.31) \\
       1 & 5 & 0 & 0 & 0.14 (0.25) & 0.06 (0.13) &	0.23 (0.28) \\
       0.001 & 50 & 0 & 0 & 0.03 (0.02) & 0.03 (0.02) & 0.06 (0.03) \\
       1 &	50 & 0 & 0 & 0.04 (0.03) & 0.03 (0.02) & 0.06 (0.04)\\
       0.001 & 1 & 0.1 & 0 & 2.71 (5.99) & 0.02 (4.07) & 8.24 (8.93)\\ 
       1 &	1 & 0.1 & 0 & 2.16 (6.04) & 0.03 (3.49) & 9.57 (10.17) \\
       0.001 & 2 & 0.1 & 0 & 0.50	(1.82) & 0.33 (0.36) & 	0.35 (1.84) \\ 
       1 & 2 & 0.1 & 0 & 0.47 (0.53) & 0.38	(0.18) & 0.28 (0.51) \\
       0.001 & 5 & 0.1 & 0 & 0.35 (0.05) & 0.35 (0.04) & 0.18 (0.02) \\
       1 &	5 & 0.1 & 0 & 0.42 (0.04) & 0.43 (0.03) & 0.21 (0.01) \\
       0.001 & 50 & 0.1 & 0 & 0.36 (0.05) &	0.35 (0.04) & 0.18 (0.01) \\
       1 & 50 & 0.1 & 0 & 0.42 (0.03) & 0.43 (0.03) & 0.21 (0.01) \\
       0.001 & 1 & 0 & 0.6 & 4.19 (8.78) & 0.02 (6.30) & 14.77 (10.31) \\
       1 & 1 & 0 & 0.6 & 4.26 (7.42) & 0.08 (5.57) & 11.58 (8.61) \\
       0.001 & 2 & 0 & 0.6 & 1.47 (4.25) & 0.05 (2.18) & 3.77 (5.08) \\
       1 & 2 & 0 & 0.6 & 1.45 (3.24) & 0.16 (1.77) & 3.28 (4.44) \\
       0.001 & 5 & 0 & 0.6 & 0.10 (0.18) & 0.04 (0.10) & 0.28 (0.92) \\
       1 & 5 & 0 & 0.6 & 0.10 (0.14) & 0.05 (0.09) & 0.19 (0.19) \\
       0.001 & 50 &	0 & 0.6 & 0.03 (0.02) &	0.03 (0.02) & 0.05 (0.04) \\
       1 & 50 &	0 & 0.6 & 0.03 (0.02) & 0.03 (0.02) & 0.06 (0.04) \\
    
       \hline
	\end{tabular}
	\end{center}
\end{table}

The results in Table \ref{tab:simres_cosk_u} illustrate some interesting patterns. At first, for approximately skew normal data ($\nu = 50$), the co-skewness of the component calculated by PP is very close to the true co-skewness. Notably this holds true regardless of the underlying covariance. This underpins that PP is suitable as a technique to identify latent mechanisms of higher order moments. Increasing the noise in the error term in \eqref{eq:lvmodel} to the level of variance of the underlying distribution does not have an impact PP's capacity to detect components with correct co-skewness. Not surprisingly, as degrees of freedom decrease, the absolute deviations increase significantly. However, it is remarkable that while the average absolute deviation increases a lot, the median absolute deviation does not. This illustrates that the method will produce erroneous estimates for individual samples generated from a skew Cauchy, not consistently for all of them. Introduction of harsh outliers does yield the expected picture for $\nu \geq 5$. That is, the classical projection index has a much larger AD compared to the runs without outliers, yet the 15\% trimmed projection index performs significantly better in that context. Slightly unexpectedly, this is not the case at all for the naturally heavy tailed distributions. Using a 15\% trimmed co-skewness as a projection index does not retrieve the correct co-skewness either with or without presence of outliers. These results for heavy tailed distributions in Table \ref{tab:simres_cosk_u} go against similar results for second-order moment statistics. It has been reported that robust methods perform better than classical ones to estimate first and second moment based statistics for Cauchy distributed data (see e.g. \cite{SDV2006}). However, in the present study, higher order moments do not fit the same picture. The absolute deviations obtained from PP with a 15\% trimmed co-skewness as projection index are a lot higher than those from PP with classical co-skewness. An explanation may be that trimming trims away not just the outliers, but also a portion of the {\em slant} of the data, leading to erroneous co-skewness estimates. It is possible that more subtly constructed robust estimators could do better, but that would require fundamental research into robust or non-parametric co-skewness estimation. 

The previous paragraph has established in which situations PP with either classical or trimmed projection indices is suitable to retrieve underlying co-skewness of a latent variable with a dependent variable. Let us now have a look at how good it is at estimating the latent variable relationship itself. The most illustrative way to do so is to have a look at the estimated loadings $\hat{\mathbf{p}}$. At first, one should consider $\mathop{\mbox{abs}}\left(\hat{\mathbf{p}}\right)$ instead of the raw loadings since this method, just like all related latent variable methods from this framework, is sign indeterminate with respect to weights, scores and loadings. Secondly, considering that the true $p_4$ loading is close to zero, the ratio $\hat{p}_4/\hat{p}_i, i \neq 4$ should be close zero as well. In Table \ref{tab:simres_pr_u}, the loading ratios (PR) $\mathop{\mbox{abs}}\left(\hat{p}_4 / \hat{p}_2\right)$ using CAPI as projection index, based on co-moments of increasing order are presented, summarized across the N=100 simulations per setting by univariate location and scale statistics. Likewise, Table \ref{tab:simres_pr_trim} summarizes results for PP---CAPI PRs based on 15\% trimmed co-moments for the exact same settings.

\begin{table}
	\caption{\label{tab:simres_pr_u} Location of the $\mathop{\mbox{abs}}\left(\hat{p}_4 / \hat{p}_2\right)$ ratio (PR) estimated by PP (CAPI) up to varying orders. Numbers between brackets represent scale estimates (standard deviation for mean, MAD for median). All simulations shown here at $\sigma_{\epsilon}=0.001$.}
	\begin{center}
	\begin{tabular}{cccccc}
		\hline
		$\nu$ & $\phi$ & cont & $\Omega_{ij}$ & 
		Mean PR, PP3(0) & 
		Median PR, PP3(0)
		\\
		\hline
		1  & 0 & 0 & 0 & 1.01 (0.42) & 0.98	(0.12)
		\\
		2 & 0 &	0 & 0 & 0.78 (1.12) & 0.30 (0.73)
		\\
		5 & 0 &	0 & 0 & 0.50 (2.00) & 0.14 (0.68) 
		\\
		50 & 0 & 0 & 0 & 0.10 (0.05) & 0.09	(0.04)
		\\
		1  & 0.1 & $\mathbf{t}$ & 0 & 1.00 (0.33) &	0.98 (0.09)
		\\
		2 & 0.1 & $\mathbf{t}$ & 0 & 1.00 (1.04) & 0.75	(0.61)
		\\
		5 & 0.1 & $\mathbf{t}$ & 0 & 0.72 (0.58) & 0.78	(0.56)
		\\ 
		50 & 0.1 & $\mathbf{t}$ & 0 & 0.70 (0.61) &	0.62 (0.59)
		\\
		1  & 0.1 & $\mathbf{X}$ & 0 & 1.10	(1.15) & 0.98 (0.31)
		\\
		2 & 0.1 & $\mathbf{X}$ & 0 & 0.94 (1.51) & 0.49	(0.80)
		\\
		5 & 0.1 & $\mathbf{X}$ & 0 & 1.25 (2.86) & 0.47	(1.39)
		\\ 
		50 & 0.1 & $\mathbf{X}$ & 0 & 0.46 (0.32) &	0.40 (0.26)  
		\\
		1  & 0 & 0 & 0.6 & 0.96	(0.06) & 0.98 (0.05) 
		\\
		2 & 0 & 0 &	0.6 & 0.29	(0.12) & 0.27 (0.08) 
		\\
		5 & 0 & 0 &	0.6 & 0.11 (0.04) &	0.12 (0.03) 
		\\
        50 & 0 & 0 & 0.6 & 0.09	(0.04) & 0.09 (0.03)
        \\
		\hline
	\end{tabular}
	\end{center}
\end{table}

\begin{table}
	\caption{\label{tab:simres_pr_trim} Location of the $\mathop{\mbox{abs}}\left(\hat{p}_4 / \hat{p}_2\right)$ ratio (PR) estimated by PP (CAPI) up to varying orders, based on 15\% trimmed moment estimates. Numbers between brackets represent scale estimates (standard deviation for mean, MAD for median). All simulations shown here at $\sigma_{\epsilon}=0.001$.}
	\begin{center}
		\begin{tabular}{cccccc}
			\hline
			$\nu$ & $\phi$ & cont & $\Omega_{ij}$ & 
			Mean PR, PP3(15) & 
			Median PR, PP3(15)
			\\
			\hline
			1  & 0 & 0 & 0 & 0.15 (0.01) & 0.15	(0.01)
			\\
			2 & 0 &	0 & 0 & 0.11 (0.02)	& 0.10 (0.02)
			\\
			5 & 0 &	0 & 0 & 0.09 (0.05) & 0.09 (0.04)
			\\
			50 & 0 & 0 & 0 & 0.09 (0.05) & 0.09	(0.04)
			\\
			1  & 0.1 & $\mathbf{t}$ & 0 & 0.24 (0.02) &	0.24 (0.02)
			\\
			2 & 0.1 & $\mathbf{t}$ & 0 & 0.12 (0.01) & 0.12	(0.01)
			\\
			5 & 0.1 & $\mathbf{t}$ & 0 & 0.09 (0.01) & 0.09	(0.00) 
			\\ 
			50 & 0.1 & $\mathbf{t}$ & 0 & 0.07 (0.00) &	0.07 (0.00)
			\\
			1  & 0.1 & $\mathbf{X}$ & 0 & 0.78 (4.49) &	0.29 (0.91)
			\\
			2 & 0.1 & $\mathbf{X}$ & 0 & 0.09 (0.03) & 0.09	(0.02)
			\\
			5 & 0.1 & $\mathbf{X}$ & 0 & 0.02 (0.01) & 0.02	(0.01)
			\\ 
			50 & 0.1 & $\mathbf{X}$ & 0 & 0.01 (0.01) &	0.01 (0.01)
			\\
			1  & 0 & 0 & 0.6 & 0.14	(0.01) & 0.14 (0.01)
			\\
			2 & 0 & 0 &	0.6 & 0.10 (0.02) &	0.09 (0.01)
			\\
			5 & 0 & 0 &	0.6 & 0.09 (0.04) &	0.09 (0.03)
			\\
			50 & 0 & 0 & 0.6 & 0.08	(0.06) & 0.08 (0.05)
			\\
			\hline
		\end{tabular}
	\end{center}
\end{table}

At first, recall that the data have been constructed according to the latent variable model \eqref{eq:lvmodel} and that the scores $\mathbf{t}$ therein have been drawn from skew $t$ distributions. Also recall that a small number of PR would be expected, since the truth is PR = 0.0015. At first, it is clear that for distributions close to the normal, fairly low PR values are observed. This holds true for both the estimates from CAPI and for estimates from CAPI based on 15\% internal trimming (henceforth named {\em robust CAPI}). In all results shown, CAPI and robust CAPI based on moments up to order three have been calculated. Results including fourth order moments are very similar in this setting, as long as the corresponding $\omega_{ij}$ are positive. They are, however, slightly less stable (not shown). Heavily tailed distributions do distort CAPI projections and so does the introduction of harsh outliers. However, the good news is that robust CAPI actually can cope with either of these disturbances, and still yields results very similar to CAPI for uncontaminated data. The only setting observed here where robust CAPI also breaks down, is for skew Cauchy distributed data with an additional 10\% of harsh outliers in $\mathbf{X}$. I conjecture that this is due to the heavily tailed Cauchy actually generating more than an additional 5\% of points far off the center, such that the amount of effective outliers exceeds 15\%. 

All results presented here have been calculated by an implementation of the method in the Python 3 package \textsf{ppdire}, which has been distributed on \textsf{PyPI} (the Python Package Index) by the Author. The simulation scripts are supplied as Supplementary Material to this manuscript.

\section{Application to financial market analysis}\label{Sec:Application}

The previous Section has illustrated that projection pursuit will find directions with co-skewness to a dependent variable very close to the true one if the data follow a latent component model. It has also been illustrated that the correct latent variable structure will be retrieved using co-skewness only as a projection index, as long as the process that generates them is not too noisy or does not generate extreme anomalies. Most financial returns data will resort under this umbrella. While the simulations are encouraging, the practical use of PP with a higher moment based projection index is much bigger when instead of an individual higher co-moment, a combination of co-moments is used. This Section will illustrate how CAPI (Equations \ref{eq:capi} and \ref{eq:capiscaled}) can come in very handy analyzing financial market data. 

For illustrative purposes, let us consider a data set of moderate dimensions, consisting of the returns of twelve shares over the last year (more specifically, from 7/1/18 through 6/30/19). Data were downloaded from Yahoo Finance as daily Open/High/Low/Close values and consecutively converted to absolute returns. The shares analyzed are listed in Table \ref{tab:shares_example}. These securities were selected to represent different types of market behaviour. Some securities are included that are typically assumed to be cyclical and correlated to the market (e.g. AMZN, retail), whereas some others are cyclical, yet more correlated to a certain commodity than to the market (e.g. CVX, energy, correlated to crude oil). Some securities were incorporated that are generally negatively correlated to the market. These correspond to precious metals mining companies and treasury bonds. The diversified nature of the shares considered, is also reflected by the betas in Table \ref{tab:shares_example}. 

\begin{table}
	\caption{\label{tab:shares_example} Securities analyzed in this example. All securities are shares listed on the New York Stock Exchange (NYSE), except for AMZN which is listed on NASDAQ. No derivatives are being considered. Source: Yahoo finance.}
	\begin{center}
		\begin{tabular}{cccc}
			\hline
			Security & Symbol & Sector & Beta\\
			\hline
			American International Group & AIG & Financial Services & 1.10 \\
			Amazon & AMZN & Consumer Cyclical & 1.73 \\ 
			Blackstone Mortgage Trust & BXMT & Real Estate & 0.41 \\
			Chevron Texaco & CVX & Energy & 0.75 \\
			Eastman Chemical & EMN & Basic Materials & 1.07 \\
			Ford Motor Co. & F & Consumer Cyclical & 0.97 \\
			Franco---Nevada & FNV & Basic Materials & -0.05 \\
			Goldman Sachs & GS & Financial Services & 1.3 \\
			Kimberly---Clark & KMB & Consumer Defensive & 0.43 \\
			Seabridge Gold & SA & Basic Materials & -0.3 \\
			Treasury Long Term Fund & TLT & Fixed Income ETF & 3.32 \\
			Wheaton Precious Metals & WPM & Basic Materials & -0.56 \\
			\hline
		\end{tabular}
	\end{center}
\end{table}

A preliminary analysis of these data shows that these data can act as a good example. There are no obvious anomalies downloaded from Yahoo finance. Some spikes in individual shares exist, but they correspond to normal market behaviour and can be interpreted. For instance, several positive and negative spikes in the Ford Motor share prices correspond to US presidential tweets on import duties for Chinese goods. These occur, and are generally interesting for analysis and thus do not need to be edited out. Besides anomalies, the returns are approximately centred about zero, and a histogram analysis leads to the conclusion that some returns are too skewed to be normally distributed (not shown). 

In what follows, CAPI will be applied by PP to these daily returns data. Moment weights in \eqref{eq:capi} are set to $\boldsymbol{\omega} = [1, .5, .5, -0.03, -0.03, -0.03]$. This means that the weights $\mathbf{w}$ reported will be more positive for securities that covary with the market, but also have positive co-skewness with the market, while having a negative co-kurtosis, even though less weight is being attributed to the last property by an order of magnitude. 

Let us at first compare PP CAPI $\hat{\mathbf{w}}$ estimates with the betas for the entire data history (1 y). Note that the Yahoo betas represented in Table \ref{tab:shares_example} are based on monthly data from a three year data history. In the present example, daily data are used for a single year, which results in different betas. The results are presented in Table \ref{tab:shares_1y_beta_capi} for CAPI accounting for moments ranging from order two until up to order four. 

% \begin{figure} 
%	\caption{ Market weight coefficients as estimated by the betas and CAPI, 5\% trimming, up to order 4.}
%	\includegraphics[width=\textwidth]{capi_weights}  
%	\label{fig:capi-weights}
% \end{figure}

\begin{table}
	\caption{\label{tab:shares_1y_beta_capi} Betas and CAPI weights for a single year of data history, returns calculated from daily OHLC data. CAPI parameters: $\boldsymbol{\omega} = [1, .5, .5, -0.03, -0.03, -0.03]$, no trimming.}
	\begin{center}
	\begin{tabular}{ccccc}
		\hline
		Symbol & beta & order 2	& order 3 & order 4 \\
		\hline
		AIG  & 0.292 & 0.299 & 0.148 & -0.035 \\
		AMZN & 0.353 & 0.473 & 0.470 & 0.412 \\
		BXMT & 0.547 & 0.281 & 0.531 & 0.492\\
		CVX  & 0.426 & 0.348 & 0.381 & 0.090\\
		EMN  & 0.408 & 0.419 & 0.431 & 0.624\\
		F	 & 0.230 & 0.259 & 0.230 & 0.244\\
		FNV	 & -0.072& -0.061& 0.015 & 0.218\\
		GS   & 0.393 & 0.401 & 0.221 & -0.158\\
		KMB	 & 0.161 & 0.139 & 0.080 & -0.116\\
		SA	 & -0.033& -0.061& 0.112 & 0.171\\
		TLT	 & -0.696& -0.234& -0.142 & 0.120 \\
		WPM	 & -0.035& -0.047& -0.047 & 0.015\\ 
		\hline
	\end{tabular}
	\end{center}	
\end{table}

At first, betas are regression coefficients and thus by default second moment based statistics. The fact that second order PP CAPI weights are very similar to the betas, shows that the method produces reasonable results. Of course, PP CAPI is most useful when applied to higher order moments. Instead of having to analyze betas, co-skewness matrices and co-kurtosis matrices and then looking for a compromise, the PP CAPI weights fulfill all of these tasks in a single numerical result. In fact, PP CAPI weights could be regarded as {\em composite betas} accounting for a specific optimization objective including higher order moments. Table \ref{tab:shares_1y_beta_capi} illustrates that as higher order moments kick in, some of the weights shift. While weights of some companies like retail giant Amazon and automotive manufacturer Ford are stable across different orders of the co-moments, some others are not at all. Notably, mining industry shares have clearly more conservative generalized beta values compared to the standard betas, when accounting for a slight resistance against co-kurtosis with the market. 

Finally, it is a good question how these weights vary over time. For what follows, PP CAPI weights have been calculated on a monthly basis, still from daily returns. The results are presented in Table \ref{tab:capi_monthly_summaries}. 

\begin{table}
	\caption{\label{tab:capi_monthly_summaries} Summary statistics of third order monthly CAPI weights for a single year of data history, returns calculated from daily OHLC data. CAPI parameters: $\boldsymbol{\omega} = [1, .5, .5]$, no trimming.}
	\begin{center}
		\begin{tabular}{ccc}
			\hline
			Symbol & mean & std   \\
			\hline
			AIG  & 0.271  & 0.148 \\
			AMZN & 0.331  & 0.197 \\
			BXMT & 0.257  & 0.128 \\
			CVX  & 0.160  & 0.128 \\
			EMN  & 0.353  & 0.123 \\
			F	 & 0.273  & 0.154 \\
			FNV	 & -0.020 &	0.269 \\
			GS	 & 0.323  &	0.202 \\
			KMB	 & 0.025  &	0.254 \\
			SA	 & -0.027 &	0.212 \\
			TLT	 & -0.100 &	0.218 \\
			WPM	 & 0.007  &	0.211 \\
			\hline
		\end{tabular}
	\end{center}	
\end{table}
Table \ref{tab:capi_monthly_summaries} shows that has been a significant variation in the third order CAPI weights in the period considered. Yet that should ot come as a surprise: between 7/2018 and 7/2019, the market paradigm fluctuated pretty intensively, caused mainly by a change of policy at the Federal reserve and by repetitive vicissitudes in the US---China trade relationship.

\section{Summary and Conclusions}\label{Sec:Conclusions}

Projection pursuit has long been known as a flexible framework to analyze multivariate data, particularly suitable if the target is to analyze data according to an optimization objective that cannot be solved analytically or would be very tedious to optimize as is numerically. Many popular statistical tools essentially fit into the PP framework, such as PCA and ICA, but for those well-known techniques, more elegant and more efficient solutions to the optimization objective exist. Therefore, in practice PP's applicability has recently mostly been restricted to the niche area of robust statistics, where it is the default way to calculate certain estimators, such as grid robust PCA \citep{CrouxFilzmoserOliveira2008} or robust continuum regression \citep{SFCV2005}.  

The present paper has shown that, owing to its flexibility, PP could see more widespread use in the future again. It has been illustrated here that PP is suitable to detect latent structures generated from models with intrinsic co-skewness. This property has been further developed into the co-moment analysis projection index (CAPI). CAPI has been illustrated to be of immediate applicability to analyze financial returns data, where the resulting weights can be seen as a generalized beta, accounting for higher order moments. The present state of the art to analyze second and higher order moments is to calculate co-moment matrices separately and then process the results according to preset logical criteria. This may both be tedious (the fourth co-moment matrix has dimensions $p \times p^3$) and may lead to inconsistent results. Using PP---CAPI produces one model consistent outcome in a single calculation, intrinsically satisfying all optimization objectives. Beyond being more convenient, PP---CAPI also offers other advantages. It is well known that co-moments are sensitive to outliers, and more so the higher the order of the co-moment. Outliers as spikes are frequent in financial market data, and depending on the objective of the analysis, one may prefer estimates robust to spikes. PP---CAPI is the first method to offer robust estimation including higher order co-moments through trimming. This, however, is not the end of the road. In contrast to classical co-skewness and co-kurtosis matrices, PP---CAPI can easily be extended to different, yet related co-moments that may be preferable in certain practical situations, such (co-)moments derived from energy statistics \citep{SzekelyRizzo2013} or ball statistics \citep{PanWXZ2019}, just to name a few.   

Whereas the results outlined in this paper are promising, they merely scratch the surface of what could be possible in applications of PP for the financial markets. Likewise, higher order moment based projection indices likely will also come in useful in areas outside financial analytics. 

\bigskip
\begin{center}
{\large\bf SUPPLEMENTARY MATERIAL}
\end{center}

\begin{description}

\item[Simulation and example code:] Python code to produce the simulations and the examples, has been uploaded as supplementary material. 

\end{description}

\bibliographystyle{/usr/local/texlive/2018/texmf-dist/bibtex/bst/natbib/plainnat.bst}

\bibliography{./bibS25.bib}

\end{document}